\documentclass[aps,prd,twocolumn,superscriptaddress,preprintnumbers,floatfix,nofootinbib,notitlepage,showkeys,showpacs]{revtex4-1}

\usepackage[utf8]{inputenc}

\usepackage{graphicx}
\usepackage{hyperref}
\usepackage{latexsym}
\usepackage{amsmath}
\usepackage{amssymb}
\usepackage{bbm}
\usepackage{ulem}
\usepackage{pdfsync}
\usepackage{epsfig}
\usepackage{epstopdf}
\usepackage{subfigure}
\usepackage{xcolor}
\usepackage{comment}
\usepackage{slashed}
\usepackage{multirow}

\definecolor{citeblue}{rgb}{0.184, 0.384, 0.431}
\definecolor{linkred}{rgb}{0.612, 0.263, 0.141}
\hypersetup{
     colorlinks=true,
     linkcolor=linkred,
     citecolor=citeblue,
     urlcolor=black
     }

\begin{document}

\title{Exploring the spectrum of stochastic gravitational-wave anisotropies with pulsar timing arrays}
\author{Gabriela Sato-Polito}
\affiliation{William H. Miller III Department of Physics and Astronomy, Johns Hopkins University,  3400 North Charles Street, Baltimore, MD 21218, United States}

\author{Marc Kamionkowski}
\affiliation{William H. Miller III Department of Physics and Astronomy, Johns Hopkins University,  3400 North Charles Street, Baltimore, MD 21218, United States}

\begin{abstract}
    Anisotropies in the nanohertz gravitational-wave background are a compelling next target for pulsar timing arrays (PTAs). Measurements or informative upper limits to the anisotropies are expected in the near future and can offer important clues about the origin of the background and the properties of the sources. Given that each source is expected (in the simplest scenario of circular inspirals) to emit at a fixed frequency, the anisotropy will most generally vary from one frequency to another. The main result presented in this work is an analytical model for the anisotropies produced by a population of inspiralling supermassive black-hole binaries (SMBHBs). This model can be immediately connected with parametrizations of the SMBHB mass function and can be easily expanded to account for new physical processes taking place within the PTA frequency band. We show that a variety of SMBHB models predict significant levels of anisotropy at the highest frequencies accessible to PTA observations and that measurements of anisotropies can offer new information regarding this population beyond the isotropic component. We also model the impact of additional dynamical effects driving the binary towards merger and show that, if these processes are relevant within the PTA band, the detectability of anisotropies relative to the isotropic background will be enhanced. Finally, we use the formalism presented in this work to predict the level anisotropy of the circular and linear polarizations of the SGWB due to the distribution of binary orientation angles with respect to the line of sight.
\end{abstract}

\maketitle

\section{Introduction}
A tantalizing detection of a common-spectrum process has now been reported across all pulsar timing arrays (PTAs) \cite{NANOGrav:2020bcs, Chen:2021rqp, Goncharov:2021oub, Antoniadis:2022pcn}, which may be the first hint of low-frequency ($\sim 1-100$ nHz) gravitational waves (GWs). Although a significant measurement of the angular cross-correlation described by the Hellings \& Downs~\cite{1983ApJ...265L..39H} curve is still lacking for a conclusive detection, the next few years of PTA observations is likely to establish the origin of this common-spectrum process. 

Supermassive black holes appear to reside at the center of most massive galaxies~\cite{Kormendy:2013dxa} and, albeit a highly uncertain process, are expected to merge following the merger of the host galaxies\cite{1980Natur.287..307B}. During their inspiral period, supermassive black-hole binaries (SMBHBs) will enter the PTA band and are the most likely sources of low-frequency gravitational waves. The PTA signal is therefore expected to be composed of the incoherent superposition of the gravitational radiation from all SMBHBs in their slow adiabatic inspiral phase (see, e.g., Ref.\cite{Sesana:2008mz}). In this regime, each system is well modelled in the Newtonian approximation, thus acting as a continuous monochromatic emitter with a frequency that is twice its orbital frequency. 


After a significant detection of an isotropic stochastic GW background (SGWB) is found, characterizing its spatial distribution will be a key scientific target to determine its origin and properties. Due to the finite number of binaries, anisotropies in the background are naturally expected from Poisson fluctuations \cite{Cornish:2013aba, Mingarelli:2013dsa}. Analytical modelling of such anisotropies has been explored for sources in the  LIGO/Virgo frequency band\cite{Jenkins:2019nks, Jenkins:2019uzp, Alonso:2020mva} in intensity and for circular polarization in Ref.~\cite{ValbusaDallArmi:2023ydl}. Furthermore, a variety of techniques to predict the impact of intensity and polarization anisotropies of the SGWB on the pulsar times-of-arrival and tools to measure such signatures from data have been developed~\cite{Mingarelli:2013dsa, Taylor:2013esa, Hotinli:2019tpc, Belgacem:2020nda, Chu:2021krj, Liu:2022skj}, with first upper limits presented using data from the European Pulsar Timing Array (EPTA)~\cite{anisotropy_uplim}. While these analyses decompose the GWB into spherical harmonics, compelling alternative techniques have been proposed to search for anisotropies, such as decomposing the GWB into eigenmaps of the Fisher matrix instead of spherical harmonics~\cite{Ali-Haimoud:2020iyz, Ali-Haimoud:2020ozu}.

The main result we present in this work is an analytical prediction for the anisotropy of the SGWB due to shot noise (given in Sec.~\ref{sec:shotnoise}), which we connect with a simple parametrization of the SMBHB mass function.  As one might expect, we show that the level of anisotropy is frequency dependent, since the number of binaries is lower and therefore the shot noise is higher at higher frequencies, scaling as $C_{\ell>0}/C_0 \propto f^{11/3}$ in the GW-dominated regime. While combining all frequency bands in a PTA washes out anisotropies, a detection or informative upper limit may be feasible within the next few years at higher frequencies. We proceed to explore a variety of signals that impact the anisotropy of the SGWB and that therefore may be constrained or measured in the near future as PTA sensitivities improve. These include {\it (i)} the SMBHB population, {\it (ii)} additional interactions driving the binary mergers, and {\it (iii)} the circular and linear polarization of the SGWB due to the distribution of inclinations angles of the binary population. 


Provided that the background is indeed produced by SMBHBs, we show that measurements of anisotropy will offer new information about the SMBHB population beyond what can be inferred from the isotropic component. In particular, while the isotropic component of the SGWB corresponds to the mean GW intensity summed over all SMBHBs ($\sim \int N h^2$), the anisotropies depend on the second moment of this distribution ($\sim \int N h^4$). Hence, measurements or upper limits to anisotropies will be an important step to maximize the information about the SMBHB population that can be retrieved from PTA data.

Our derivation also provides a simple framework to predict anisotropies while relaxing some of the assumptions that are typically made in the standard scenario of GW emission within the PTA band. As a corollary of our results, we show how the frequency dependence of anisotropies is modified by additional physical processes driving the binary inspiral, such as stellar scattering and interactions with circumbinary gas. These processes might accelerate the evolution of the binary at the lowest frequencies accessible to PTA observations. In this scenario, the GW spectrum is attenuated, but we show that the relative amplitude of the anisotropies are enhanced.

Finally, we use this formalism to compute the expected level of linear and circular polarizations of the SGWB. Each SMBHB is individually polarized as a function of the orientation of the orbit with respect to the line of sight. While the expectation value of the isotropic component of the linear and circular polarizations are zero, the anisotropies are not, since fluctuations in the numbers of sources lead to fluctuations on the polarizations at different sky locations. We show that, given a prediction for the level of anisotropies in the intensity, the circular polarization is expected to comparable to this value (as shown by Ref.~\cite{ValbusaDallArmi:2023ydl}), and the linear polarizations around $3\%$.

\begin{figure*}
    \centering
    \includegraphics[width=\textwidth]{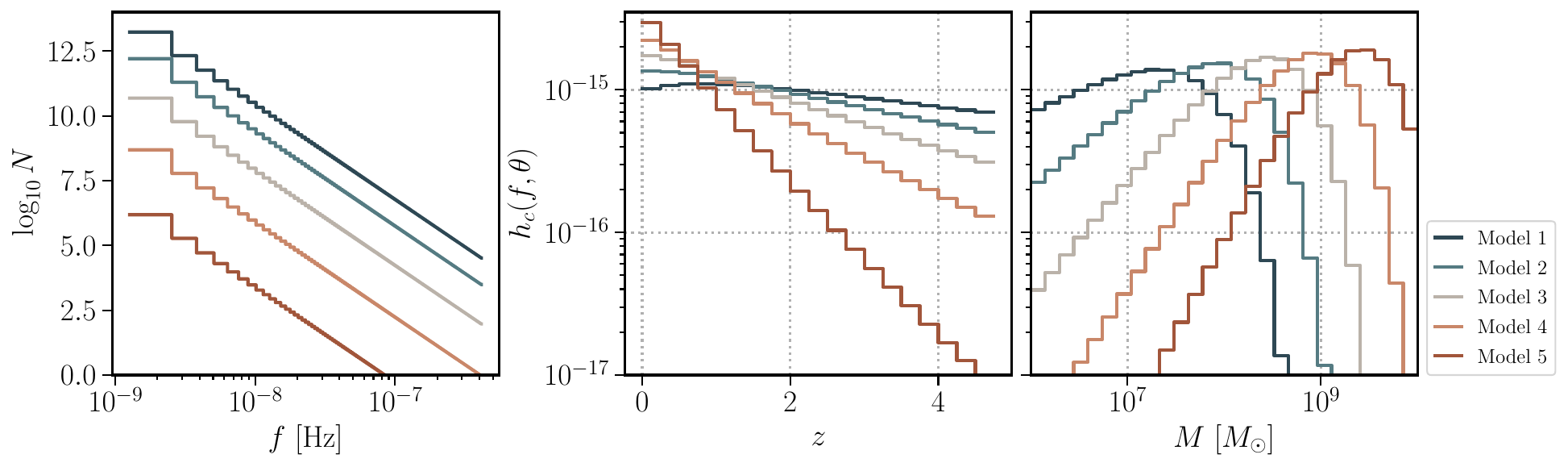}
    \caption{Number of SMBHBs emitting in each frequency bin of a PTA and the contributions of each redshift and chirp mass bin to the total characteristic strain. In the leftmost plot, we assume that the frequency bin width is fixed by the inverse of the observing time, which we assume to be 20 yr. Each color corresponds to a different model for the SMBHB mass function, specified in Table~\ref{tab:models}.}
    \label{fig:hc_N_models}
\end{figure*}

This paper is organized as follows. We discuss the model for the SGWB produced by a population of inspiralling SMBHBs in Sec.~\ref{sec:SGWB}, which has been used to compute the isotropic component. We extend this formalism to compute the anisotropies of the SGWB in Sec.~\ref{sec:shotnoise} due to shot noise. We present our results in Sec.~\ref{sec:results}, exploring the dependence of the anisotropies on the SMBHB population, the physical processes driving the binary inspiral, and the polarization of the SGWB. We conclude in Sec.~\ref{sec:conclusion}.

\section{SGWB model}\label{sec:SGWB}

A gravitational-wave background is composed from the sum of all SMBHBs emitting at a given observed frequency. Under the assumption that the evolution of the systems in the PTA band are stationary and dominated by GW emission, a model for the background only requires an additional model for the number density of SMBHBs per chirp mass and redshift. In this scenario, the characteristic strain at a frequency $f$ is given by \cite{Phinney:2001di, Sesana:2008mz}

\begin{equation}
\begin{split}
    h^2_c(f) =& \frac{4G}{\pi c^2 f^2}\int d\log_{10}\mathcal{M} \int \frac{dz}{1+z} \\ &\times \frac{dn}{dzd\log_{10}\mathcal{M}} \left. \left(\frac{d E_{\rm gw}}{d\log f_r}\right)\right|_{f_r=(1+z)f},
\label{eq:h2c}
\end{split}
\end{equation}
where $dn/dzd\log_{10}\mathcal{M}$ is the number density of sources per redshift and logarithmic chirp mass bin, and $dE_{\rm gw}/d\log f_r$, the rest-frame energy spectrum generated by individual sources. The relevant redshift and chirp mass ranges in the integrals above are $0\leq z \leq 5$ and $10^6 \leq \mathcal{M}/M_{\odot} \leq 10^{11}$.

We adopt a simple model for the SMBHB number density, described in Ref.~\cite{Middleton:2015oda},
\begin{equation}
\begin{split}
    \frac{dn}{dz d\log_{10}\mathcal{M}} =&\ \dot{n}_0 \left[ \left(\frac{\mathcal{M}}{10^7 M_{\odot}}\right)^{-\alpha} e^{-\mathcal{M}/\mathcal{M}_*} \right] \\
    &\ \times \left[(1+z)^{\beta} e^{-z/z_0}\right] \frac{dt_r}{dz},
\label{eq:dndzdm}
\end{split}
\end{equation}
where $t_r$ is the coordinate time in the source rest frame, $\dot{n}_0$ is the normalized merger rate, the parameters $\beta$ and $z_0$ describe the redshift distribution and $\alpha$ and $\mathcal{M}_*$ the chirp mass distribution.

It will be convenient to describe the total number of binaries $N$ in a spherical shell of thickness $dz$, per logarithmic mass bin, and emitting at a frequency $f_r$. The number density of binaries in Eq.~\ref{eq:dndzdm} can be rewritten as
\begin{equation}
    \frac{dn}{dz d\log_{10}\mathcal{M}} = \frac{dN}{dz d\log_{10}\mathcal{M} d\log f_r}\frac{d\log f_r}{dt_r} \frac{dt_r}{dz}\frac{dz}{dV_c}.
\end{equation}
The change in the emitted frequency over time is assumed to be only due to the shrinking of the orbital radius from GW emission and is therefore given by
\begin{equation}
    \frac{d\log f_r}{dt_r} = \frac{96}{5}\pi^{8/3}\left(\frac{G\mathcal{M}}{c^3}\right)^{5/3}f_r^{8/3},
    \label{eq:dlogf_dt}
\end{equation}
and $V_c$ is the comoving volume, and $(dt_r/dz)(dz/dV_c) = [4\pi c (1+z)D_A^2]^{-1}$. Under these assumptions, the energy spectrum is given by
\begin{equation}
    \frac{d E_{\rm gw}}{d\log f_r}= \frac{1}{3G} \left(G\mathcal{M}\right)^{5/3}(\pi f_r)^{2/3},
\end{equation}
and the characteristic strain in Eq.~\ref{eq:h2c} can therefore be rewritten as
\begin{equation}
    h^2_c(f) = \int d\log_{10} \mathcal{M} \int dz \frac{dN}{dz d\log_{10}\mathcal{M} d \log f} h^2(z,\mathcal{M}, f),
\label{eq:h2c_N}
\end{equation}
where $h^2$ is the time and orientation averaged strain from binaries within an infinitesimal bin in redshift, mass, and frequency, such that
\begin{equation}
    h^2(z, \mathcal{M}, f) = \frac{32 \pi^{4/3}}{5c^8}\frac{(1+z)^{10/3}}{d^2_L(z)} \left(G \mathcal{M}\right)^{10/3} f^{4/3},
    \label{eq:h2}
\end{equation}
where $d_L$ is the physical (non-comoving) luminosity distance and all of the frequencies have been expressed in terms of observed (not rest-frame) frequencies. We note that Eq.~\ref{eq:h2} is equivalent to $h^2 = \langle h^2_{+} + h^2_{\times}\rangle_{t, \hat{\Omega}_{\iota}}$, which can be shown by using that the total power radiated in GWs is given by
\begin{equation}
    \frac{dE_{\rm gw}}{dt } = \frac{c^3 d^2_L(z)}{16 \pi G (1+z)^2} \int d\hat{\Omega}_{\iota} \langle \dot{h}^2_{+} + \dot{h}^2_{\times} \rangle_t,
    \label{eq:dEdt}
\end{equation}
and that $\langle \dot{h}^2_{+} + \dot{h}^2_{\times} \rangle_t = 4\pi^2 f_r^2 \langle h^2_{+} + h^2_{\times} \rangle$.

We select a range of models that will span lower to higher levels of anisotropy, but that are normalized to have the same characteristic strain spectrum $h_c(f)$. The parameters of Eq.~\ref{eq:dndzdm} for each model are given in the table below. We show in Figure~\ref{fig:hc_N_models} the mean number of SMBHBs per frequency bin and the contribution to the characteristic strain from each redshift and chirp mass bin for each of the models considered. The models range from being dominated by a large number of low-mass sources, with a broad redshift distribution (blue), to those dominated by a small number of very massive and loud nearby binaries (red). In each model, we choose the value of $\dot{n}_0$ such that the total $h_c(f)$ is fixed to a value consistent with the common-spectrum process recently found in recent PTA observations, with a value of $2\times 10^{-15}$ at a reference frequency of $f=1$yr$^{-1}$.

\begin{table}[h]
    \centering
    \begin{tabular}{|c|c|c|c|c|}\hline
      Model & $\alpha$ & $\mathcal{M}_*$ [$M_{\odot}$] & $\beta$ & $z_0$ \\ \hline \hline
      (1) & 1 & $3.2\times 10^7$ & 3 & 3 \\ \hline
      (2) & 0.5 & $7.5\times 10^7$ & 2.5 & 2.4 \\ \hline
      (3) & 0 & $1.8\times 10^8$ & 2 & 1.8 \\ \hline
      (4) & -0.5 & $4.2\times 10^8$ & 1.5 & 1.1 \\ \hline  
      (5) & -1 & $10^9$ & 1 & 0.5 \\ \hline  
    \end{tabular}
    \caption{Parameters of each SMBHB mass function model used in Figs.~\ref{fig:hc_N_models} and \ref{fig:cls}.}
    \label{tab:models}
\end{table}

As has been pointed out in Ref.~\cite{Sesana:2008mz}, the model described above overpredicts the characteristic strain at high frequencies. This is due to the fact that the integrals in Eq.~\ref{eq:h2c_N} sum over bins in mass and redshift where the mean number of binaries is less than 1 for certain frequecies. Hence, for SMBHBs in such bins, any realization of this distribution will most likely contain zero sources. In other words, while Eqs.~\ref{eq:h2c} and \ref{eq:h2c_N} accurately predict the mean characteristic strain, the median can potentially be much lower. This can be accounted for by placing an upper limit in the SMBHB mass integral $\bar{\mathcal{M}}$ in either equations, where $\bar{\mathcal{M}}$ for each frequency bin is found by imposing that
\begin{equation}
\int_{\bar{\mathcal{M}}}^{\infty}d\mathcal{M} \int^{f+\Delta f/2}_{f-\Delta f/2} df \int dz \frac{dN}{dzd\mathcal{M}df} = 1.
\end{equation}

\begin{figure*}[t]
    \centering
    \includegraphics[width=\textwidth]{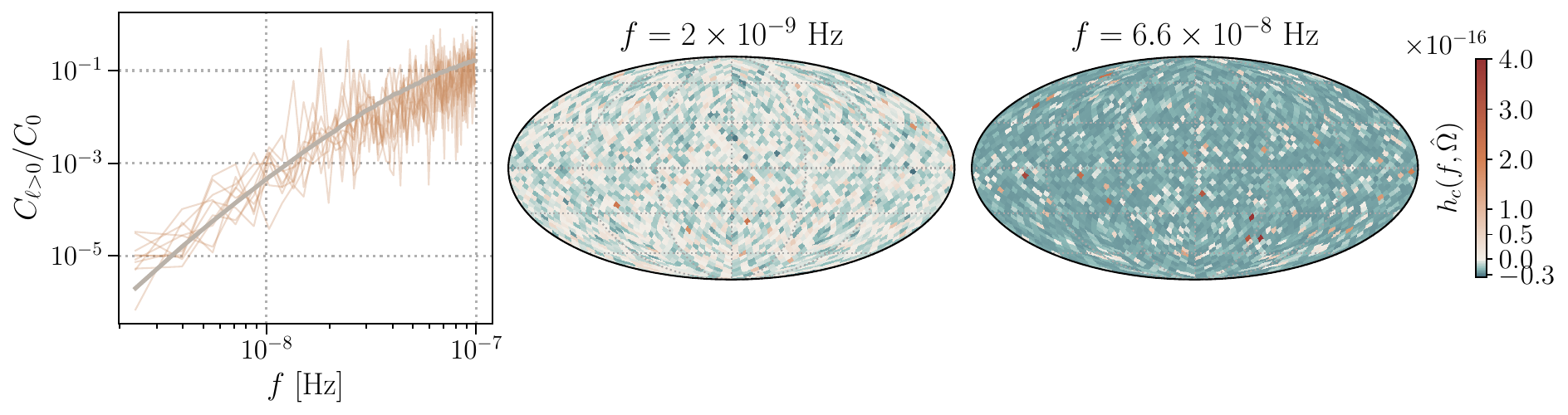}
    \caption{Anisotropy of the SGWB from realizations of a Poisson distribution of sources. The panel on the left shows the theoretically predicted anisotropy from Eq.~\ref{eq:c_ell} (thick grey line) and the values obtained from measurements of a simulated sky map (thick orange lines). The simulations consist of sampling a number of SMBHBs in each mass, redshift, and frequency bin from a Poisson distribution, uniformly distributing them on the sky, and measuring the spherical harmonic coefficients of this map. The two panels on the right show maps of the mean subtracted characteristic strain of two different frequency bins of one realization.}
    \label{fig:map}
\end{figure*}

\section{Anisotropic SGWB}\label{sec:shotnoise}
Eq.~\ref{eq:h2c} corresponds to the total characteristic strain of gravitational waves and is therefore understood as a sum across all sources over the entire sky. That is,
\begin{equation}
    h_c^2(f) = \int d\hat{\Omega}\ h_c^2(f, \hat{\Omega}),
\end{equation}
where the characteristic strain $h_c^2(f, \hat{\Omega})$ arriving from a given angular position $\hat{\Omega}$ depends on the number of sources at that location on the sky, which we assume to be isotropically distributed (thereby neglecting the effect of clustering). Hence,
\begin{equation}
    h_c^2(f, \hat{\Omega}) = \int d\vec{\theta}\ \frac{dN}{d\vec{\theta} d\log f d\hat{\Omega}}(\hat{\Omega}) h^2(f, \vec{\theta}),
    \label{eq:h2c_Om}
\end{equation}
where the variables that characterize the binary (here, $z$ and $\log_{10}\mathcal{M}$) have been abbreviated as $\vec{\theta}$.

The mean number of SMBHB is assumed to be constant throughout space, but any realization of the sky can be modelled as a finite number of discrete sources, sampled from a Poisson distribution. Thus, the discreteness of the number of sources induces an inherent fluctuation on the amplitude of GW emissions as a function of position, which is expected to be the primary source of anisotropy in the SGWB. 

In analogy with shot-noise calculations in the context large-scale structure, we may divide the sky into a grid of discrete cells, where $\hat{\Omega}_i$ denotes the angular position of the center of each cell $i$, which contains either $N_i = 0$ or $1$ objects emitting at a frequency $f$ within $[\log f, \log f+d\log f]$. A useful property that follows from this is that $N_i^2 = N_i$. Since the frequency bin is determined by the total observing time of the pulsars, it will be convenient to consider finite frequency bin widths $\Delta \log f$, instead of infinitesimal ones.

The mean number of binaries with parameters $\vec{\theta}$ within a given cell $i$ is 
\begin{equation}
\begin{split}
    \langle N_i \rangle (f, \vec{\theta})  =& \frac{f}{4\pi \Delta f} \int \limits_{\Delta \log f}d\log \tilde{f} \frac{dN}{d\vec{\theta}d\log \tilde{f}}  d\vec{\theta} \\
    \equiv& \frac{f}{4\pi \Delta f} \frac{dN_{\Delta f}}{d\vec{\theta}}d\vec{\theta},
\end{split}
\end{equation}
where the second line defines the mean number of binaries in a finite-width frequency bin, and infinitesimal bins of properties $\vec{\theta}$. The GW amplitude from such SMBHBs is therefore given by
\begin{equation}
h^2_c(f, \vec{\theta}, \hat{\Omega}_i) d\vec{\theta} = N_i h^2 d\vec{\theta},
\end{equation}
where $N_i$ and $h^2$ are both functions of frequency, and additional parameters $\vec{\theta}$. The correlation between the GW amplitudes emitted from positions $\hat{\Omega}_i$ and $\hat{\Omega}_j$ from binaries with properties $\vec{\theta}$, $\vec{\theta}'$ and frequencies $f$, $f'$ is therefore
\begin{equation}
\langle h^2_c(f, \vec{\theta}, \hat{\Omega}_i) h^2_c(f', \vec{\theta}', \hat{\Omega}_j) \rangle d\vec{\theta} d\vec{\theta}' = \langle N_i N_j \rangle h^2 h'^2 d\vec{\theta} d\vec{\theta}'.
\label{eq:h2c_cov}
\end{equation}
When $i \neq j$, $N_i$ and $N_j$ are uncorrelated, whereas when $i = j$, we may use the fact that there can only be 0 or 1 SMBHBs in the cell. Hence,
\begin{equation}
\begin{split}
    \langle N_i N_j \rangle_{i\neq j} =& \langle N_i \rangle \langle N_j \rangle \\
    \langle N_i N_j \rangle_{i=j} =& \langle N_i \rangle \delta(\vec{\theta}, \vec{\theta}')
\end{split}
\end{equation}
Substituting this result into Eq.~\ref{eq:h2c_cov} and integrating over all binary parameters $\vec{\theta}$ and $\vec{\theta}'$, we find
\begin{equation}
\begin{split}
    \langle & h^2_c(f, \hat{\Omega}_i) h^2_c(f', \hat{\Omega}_j) \rangle = \frac{\delta(f-f')}{(4\pi)^2}  \left(\frac{f}{\Delta f}\right)^2 \\ &\times \Bigg[\left(\int d\vec{\theta}\ \frac{dN_{\Delta f}}{d\vec{\theta}} h^2(f,\vec{\theta})\right)^2 + \delta_{ij}\int d\vec{\theta}\ \frac{dN_{\Delta f}}{d\vec{\theta}} h^4(f,\vec{\theta})\Bigg].
    \label{eq:corrfunc}
\end{split}
\end{equation}

By decomposing $h^2_c(f, \hat{\Omega})$ into spherical harmonics it is straightforward to show that the predicted angular power spectrum is given by
\begin{equation}
\begin{split}
    \langle a^*_{\ell m}(f) a_{\ell' m'}(f') \rangle = \delta_{\ell \ell'}\delta_{m m'} \delta(f-f') C_{\ell}(f)& \\
    = \delta_{\ell 0} \delta_{m 0}\delta(f-f') \left(\frac{f}{4\pi \Delta f}\int d\vec{\theta}\ \frac{dN_{\Delta f}}{d\vec{\theta}} h^2(f,\vec{\theta})\right)^2& \\
    + \delta_{\ell \ell'} \delta_{m m'} \delta(f-f')\left(\frac{f}{4\pi \Delta f}\right)^2 \int d\vec{\theta}\ \frac{dN_{\Delta f}}{d\vec{\theta}} h^4(f,\vec{\theta})&.
    \label{eq:c_ell}
\end{split}
\end{equation}

We compare in Fig.~\ref{fig:map} the result from the analytical prediction given in Eq.~\ref{eq:c_ell} with the $C_{\ell}$ measured from a simulated SGWB map. In order to simulate the GW sky, we compute the predicted number of binaries in each bin in redshift, mass, and frequency, and obtain the number of binaries for each realization by sampling a Poisson distribution with a mean given by this value. Angular positions on the sky for each binary are then uniformly sampled (in $\cos \theta$ and $\phi$) and the GW intensity of each binary, computed using Eq.~\ref{eq:h2}, is then added to the map. In the leftmost panel of Fig.~\ref{fig:map}, we find a good agreement between our theoretical prediction and the measurements from the simulation. The two panels on the right show the GWB map for one of the realizations at different frequencies, illustrating the higher level of anisotropy in the higher frequency compared to the lower. Note that the mean intensity is subtracted from the map.

\begin{figure*}[t]
    \centering
    \includegraphics[width=0.85\textwidth]{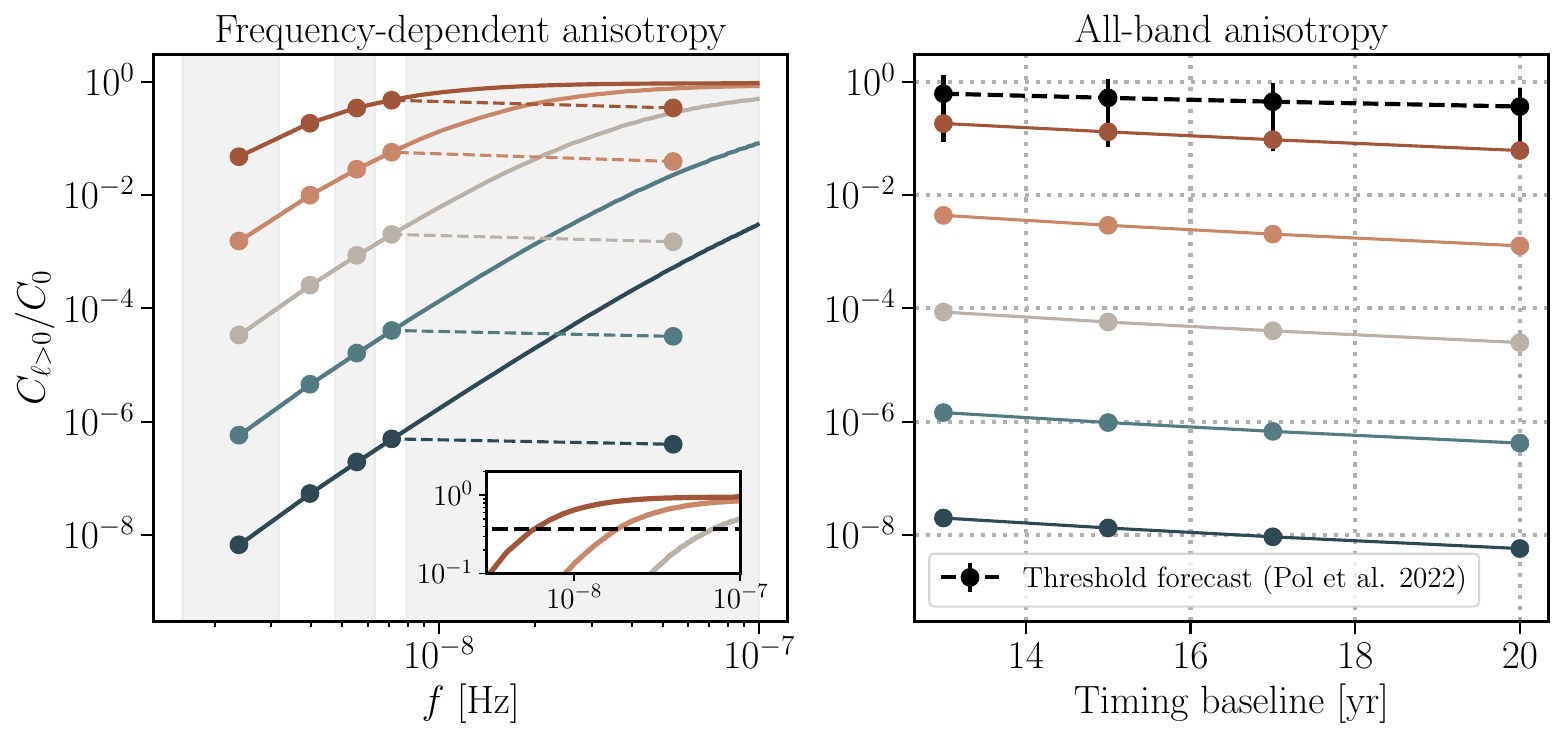}
    \caption{Angular power spectrum of the SGWB relative to the isotropic component for each of the models listed in Table~\ref{tab:models}, adopting the same legend as Fig.~\ref{fig:hc_N_models}. The frequency-dependent anisotropy is shown in the left panel, where the solid lines correspond to frequency bins set by the total observing time and the dashed lines with dots correspond to one of the frequency parametrizations adopted in Ref.~\cite{anisotropy_uplim}. The panel on the right shows the anisotropy across all frequencies, compared to the threshold for a $3\sigma$ detection of the $\ell=1$ multipole forecasted in Ref.~\cite{Pol:2022sjn}.}
    \label{fig:cls}
\end{figure*}

\section{Results and Discussion}\label{sec:results}
The derivation presented in Sec.~\ref{sec:shotnoise} depends on a number of astrophysical quantities and assumptions, while being sufficiently simple so that they can be varied or expanded upon. In this section, we apply this model of the SGWB for different models of the SMBHB mass function and consider the presence of additional dynamical interactions driving the binary inspiral, thereby investigating what a future detection or upper limit to anisotropies can teach us about such phenomena. We also quantify the expected level of polarization of the SGWB, which can serve as an indicator for the orientation of the binaries with respect to the line of sight, even if individual sources cannot be detected, as well as provide a consistency check given a detection of intensity.
\subsection{SMBHB mass function}
Using Eq.~\ref{eq:c_ell}, we compute the relative level of anisotropy $C_{\ell > 0}/C_{0}$ for each of the models shown in Fig.~\ref{fig:hc_N_models}, which we show in Fig.~\ref{fig:cls}. We emphasize that, since all models are defined to have identical characteristic strain values, {\it any detection or upper limit to anisotropy will offer new information regarding the SMBHB mass function.} We also show precisely what that new information is: while the monopole depends primarily on the first moment squared of the GW intensity $\sim (\int d\vec{\theta} \frac{dN}{d\theta} h^2)^2$, anisotropies depend on the second moment of the GW intensity $\sim \int d\vec{\theta} \frac{dN}{d\theta} h^4$. 

The left plot in Fig.~\ref{fig:cls} shows our prediction for the anisotropy in each frequency bin, assuming a total observing time of $T_{\rm obs}=20$yr and a maximum frequency of $10^{-7}$Hz. The solid line corresponds to the result using the smallest (linear) frequency bins, set by $f_{\rm min} = 1/T_{\rm obs}$. In practice, however, PTAs lose sensitivity at higher frequencies and bins are usually combined to mitigate this. The dashed line with dots marking the bin center corresponds to one of the approaches to frequency binning adopted in Ref.~\cite{anisotropy_uplim}, which they refer to as frequency-dependent anisotropy parametrization {\it(ii)}. In this case, the 4 lowest bins are unaltered, while the remaining are combined into a single bin, illustrating the degree to which this choice of binning will wash out anisotropies. While a variety of models predict significant levels of anisotropy within the PTA band, the sensitivity to higher frequencies and a careful consideration of the optimal frequency binning to maximize signal-to-noise may be crucial to achieve a detection.

In the panel on the right of Fig.~\ref{fig:cls}, we compute the anisotropy for all frequency bands combined and compare it with the sensitivity expected from NANOGrav in the near future, computed in Ref.~\cite{Pol:2022sjn}. The black curve corresponds to the minimum level anisotropy at $\ell=1$ required for a $3\sigma$ detection, as a function of the timing baseline. While an all-band anisotropy parametrization is often considered in the literature, Fig.~\ref{fig:cls} indicates that such a parametrization makes a detection extremely unlikely in the near future. We also compare the threshold level of anisotropy from Ref.~\cite{Pol:2022sjn}, considering the 20-yr baseline, to the frequency-dependent anisotropy in the inset plot in the left panel. We stress that this comparison is {\it not} apples-to-apples, since splitting up the data into frequency bins will lead to a loss in sensitivity, but show this as a rough indication of which models predict significant levels of anisotropy at each frequency.

\subsection{GW spectrum}
The merger of SMBHBs is expected to occur after a sequence of dynamical interactions following the merger of the host galaxies. During the initial stages of the merger, dynamical friction will drag the SMBHs towards the center of the merger remnant, which can produce a bound system when this mechanism is efficient. If  embedded in a star- or gas-rich environment, a variety of dynamical effects can drive then binary further together, such as the slingshot mechanism in three-body interactions with stars and angular momentum transfer between the binary and the disk. Such dynamical effects can potentially leave imprints on PTA observations. Ref.~\cite{2011MNRAS.411.1467K}, for instance, showed that interactions between the SMBHBs and the surrounding gas can efficiently shrink the binary separations and that the lowest frequencies in the PTA band may fall near the transition between the gas- and GW-driven regimes.

\begin{figure}[h]
    \centering
    \includegraphics[width=0.43\textwidth]{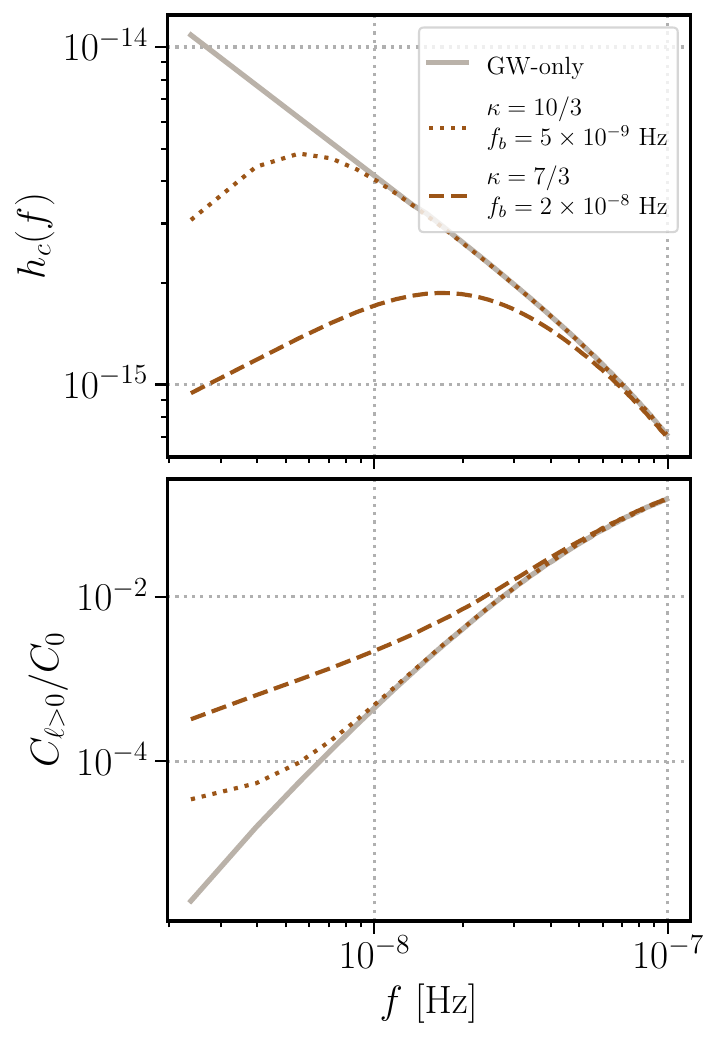}
    \caption{Impact on the GW spectrum and anisotropies of additional mechanisms driving the binary inspiral. The solid grey line corresponds to model 3 (same as shown in Figs.~\ref{fig:hc_N_models} and \ref{fig:cls}), where the SMBHBs are assumed to shrink only due to GW emission. The dotted line corresponds to a power law akin to the effect expected from stellar scattering and the dashed line corresponds to a model that includes interactions with a gaseous circumbinary disc.}
    \label{fig:GWspec}
\end{figure}



In any of these scenarios, binaries shrink faster due to the additional mechanism, which reduces the number of sources at lower frequencies relative to the GW-only case. While this attenuates the GW spectrum, it will enhance the expected level of anisotropies. Our results offer a straightforward model to self-consistently predict the impact of such effects on both the spectrum and anisotropies of the SGWB. Following a similar approach as Ref.~\cite{Sampson:2015ada}, we can see from Eq.~\ref{eq:h2c_N} that the spectrum is proportional to
\begin{equation}
    h_c^2(f) \propto \frac{dN}{d\log f} h^2(f) \propto \frac{dt}{d\log f} \left(\frac{1}{f^2}\frac{dE}{dt}\right) \propto f^{-4/3},
\end{equation}
where the last equality corresponds to the result for the standard GW-driven scenario. Applying the same analysis to the predicted level of anisotropies leads us to conclude that
\begin{equation}
    \frac{C_{\ell}}{C_0}(f) \propto \frac{1}{1+ \frac{\left(\frac{dt}{d\log f} h^2 \right)^2}{f\frac{dt}{d\log f} h^4}} \propto \frac{1}{1+(f/f_*)^{-11/3}},
\end{equation}
where $f_*$ is a critical frequency that depends on the SMBH mass function. The power-law of $\propto f^{11/3}$ is precisely what we observe at low frequencies in Fig.~\ref{fig:cls}.

Different dynamical effects that can alter the binary inspiral can be accounted for by including additional contributions (labelled $i$) to the frequency evolution as \cite{Sampson:2015ada}
\begin{equation}
    \frac{d\log f}{dt} = \left(\frac{d\log f}{dt}\right)_{\rm gw} + \sum_{i} \left(\frac{d\log f}{dt}\right)_i.
\end{equation}
Starting from Eqs.~\ref{eq:h2c} or \ref{eq:h2c_N} as a model for the GW-only spectrum, which we now rename to $h^2_{c, {\rm gw}}$, any new mechanism that drives the inspiral can be generically included as
\begin{equation}
    h^2_c(f) = \left(\frac{1}{1+\left(f_b/f\right)^{\kappa}}\right) h^2_{c, {\rm gw}}(f),
\end{equation}
where 
\begin{equation}
    \left(\frac{f_b}{f}\right)^{\kappa} = \left(\frac{d\log f}{dt}\right)_i \left(\frac{dt}{d\log f}\right)_{\rm gw},
\end{equation}
and therefore the "bend" frequency $f_b$ is the characteristic frequency in which one mechanisms supersedes the other. As discussed in Refs.~\cite{Sampson:2015ada, 2011MNRAS.411.1467K}, $\kappa = 10/3$ is a representative value for stellar scattering and $\kappa = 7/3$ for interactions with the circumbinary gas. Additional dynamical effects therefore modify the anisotropies by
\begin{equation}
    \frac{C_{\ell}}{C_0}(f) \propto \frac{1}{1+ \frac{(f/f_*)^{-11/3}}{1+\left(f_b/f\right)^{\kappa}}}.
\end{equation}

We show this result in Figure~\ref{fig:GWspec} for a few characteristic values of $f_b$ and $\kappa$. We find agreement with the qualitative understanding that, since the presence of additional dynamical effects will accelerate the binary evolution, the reduction in the number of sources will boost anisotropies at lower frequencies. This is consistent with the result found by Ref.~\cite{2011MNRAS.411.1467K}, which found an increase in the detection significance for individual sources. We therefore conclude that if any additional dynamical mechanisms are at play within the PTA band, the detectability of anisotropies in the SGWB relative to the isotropic component will be enhaced.

One potentially significant effect that has been neglected in this study is the role of eccentricity. It is likely that the presence of the aforementioned dynamical effects would drive an increase in the eccentricity of the binary, leading it to radiate in a spectrum of harmonics of the orbital frequency~\cite{Taylor:2015kpa, Chen:2016zyo}. Since the binaries would no longer be monochromatic, the GW emission at each frequency would receive contributions from a larger number of sources, which would both reduce the level of anisotropy and enhance the correlation in the anisotropy between different frequency bands. That being said, GW emission will circularize the binary and the level of residual eccentricity is unknown. We leave further investigation of this effect for future work.

\subsection{Linear and circular polarization of GWs}
Finally, we apply the formalism presented in Sec.~\ref{sec:shotnoise} to compute the amplitude of linear and circular polarization components of the SGWB, relative to its intensity. Similarly to electromagnetic waves, we may define linear and circular polarization of gravitational waves through the phase difference in the oscillations of $h_+$ and $h_{\times}$  (see, e.g., \cite{Kato:2015bye, Hotinli:2019tpc, Belgacem:2020nda, Sato-Polito:2021efu}). The gravitational radiation produced by individual SMBHBs is inherently polarized, as a function of its orientation with respect to the line of sight, since 
\begin{equation}
\begin{split}
    h_+(t) =& A \bigg[ \frac{1+\cos^2\iota}{2}\cos\Phi(t)\cos2\psi - \cos\iota\sin \Phi(t)\sin2\psi \bigg]\\
    \equiv& \mathrm{Re}\big[A e^{i\Phi(t)} \mathcal{E}_{+}\big], \\
    h_{\times}(t) =& A \bigg[\cos\iota\sin\Phi(t)\cos2\psi + \frac{1+\cos^2\iota}{2}\cos\Phi(t)\sin2\psi \bigg] \\
    \equiv& \mathrm{Re}\big[A e^{i\Phi(t)}\mathcal{E}_{\times} \big],
    \label{eq:h_+x}
\end{split}
\end{equation}
where $\Phi(t)= \phi_0 + 2\pi f$, $\phi_0$ is the initial phase, $\psi$ is the gravitational wave polarization angle, $\iota$ is the angle between the plane of the binary orbit and the line of sight, and $A=\frac{4(1+z)^{5/3}}{d_L}\left(G\mathcal{M}/c^2\right)^{5/3} (\pi f/c)^{2/3}$ is the GW amplitude. 

The model presented in Secs.~\ref{sec:SGWB} and \ref{sec:shotnoise} can be extended to include a description of the polarization of the SGWB produced by SMBHBs. Starting again from Eq.~\ref{eq:h2c_N}, we note that the strain average $h^2$ given in Eq.~\ref{eq:h2} was defined in terms of the intensity of the GWs $\langle h^2_{+} + h^2_{\times}\rangle$. We expand this definition to describe the circular and linear polarizations, such that $h^2_{X} = A^2 g_{X}(\iota)$ where $X=I, Q, U, V$, and
\begin{equation}
\begin{split}
    g_I \equiv& |\mathcal{E}_{+}|^2 + |\mathcal{E}_{\times}|^2 = \frac{1}{4}\left(1+\cos^2 \iota\right)^2 + \cos^2\iota, \\
    g_Q \equiv& |\mathcal{E}_{+}|^2 - |\mathcal{E}_{\times}|^2 = \frac{1}{4}\cos 4\psi \sin^4\iota, \\
    g_U \equiv& 2\mathrm{Re}\left[\mathcal{E}_{+}\mathcal{E}^*_{\times}\right] = \frac{1}{4}\sin 4\psi \sin^4\iota, \\
    g_V \equiv& -2\mathrm{Im}\left[\mathcal{E}_{+}\mathcal{E}^*_{\times}\right] = \frac{1}{2}\cos\iota(1+\cos^2\iota).
    \label{eq:stokes}
\end{split}
\end{equation}

In principle, the linear polarization field of the SGWB must be described in terms of spin-4 spherical harmonics \cite{Conneely:2018wis}. Since we only wish to estimate the amplitude of the linear polarization anisotropies, we avoid such complications by defining the scalar $h^2_{L} = \sqrt{h^4_{Q} + h^4_{U}}$. The angular dependence of $L$ is then given by
\begin{equation}
    g_{L}(\iota) = \frac{\sin^4 \iota}{4}.
\end{equation}

Each binary is now additionally characterized by its orientation angle with respect to the line of sight $\hat{\Omega}_{\iota}$. Using Eq.~\ref{eq:h2c_Om}, the GW amplitudes for intensity and the polarization content of a population of SMBHBs is therefore given by
\begin{equation}
    h^2_{c,X}(f, \hat{\Omega}) = \int d\vec{\theta} \int d\hat{\Omega}_{\iota} \frac{dN}{d\vec{\theta}d\log f d\hat{\Omega} d\hat{\Omega}_{\iota}} h^2_X(f, \hat{\Omega}_{\iota}), 
\end{equation}
where $X=I, V, L$ and the orientation of the binaries are assumed to be isotropically distributed. The equation above recovers the result from Sec.~\ref{sec:SGWB} for the case of intensities, and predicts a mean value of zero for circular polarization and each component of linear polarization (although not for $L$).

 Following through the same calculation presented in Sec.~\ref{sec:shotnoise}, we find from Eq.~\ref{eq:corrfunc} that the variance of the intensity and polarization fields are proportional to
\begin{equation}
    \begin{split}
        \langle h_{c,I}^4 \rangle - \langle h_{c,I}^2 \rangle^2 \propto& \int\frac{d\Omega_{\iota}}{4\pi} g^2_I(\iota) = \frac{284}{315},\\
        \langle h_{c,V}^4 \rangle - \langle h_{c,V}^2 \rangle^2 \propto& \int\frac{d\Omega_{\iota}}{4\pi} g^2_V(\iota) = \frac{92}{105},\\
        \langle h_{c,L}^4 \rangle - \langle h_{c,L}^2 \rangle^2 \propto& \int\frac{d\Omega_{\iota}}{4\pi} g^2_{L}(\iota) = \frac{8}{315}.\\
    \end{split}
\end{equation}
Hence, for any model prediction of the SGWB, the circular polarization anisotropy should be comparable to the intensity, and the linear polarization should be around 3\%, with the variance of each component $Q$ and $U$ expected to be half that of $L$. 

\section{Conclusions}\label{sec:conclusion}
After a detection of the isotropic SGWB is achieved, anisotropies will become a compelling next target for PTA measurements. A detection or upper limit on the anisotropies can help establish whether their physical origin is cosmological or astrophysical, and characterize the properties of the sources. The primary result presented in this work is a simple analytical model for the anisotropies of a background produced by inspiralling SMBHBs that can be immediately connected with the astrophysical properties of the sources and modified to account for new physical processes occurring within the PTA frequency band.

Our results show that anisotropies offer new information regarding the SMBHB population beyond the isotropic component, which can break degeneracies present in the current analysis (see, e.g., Ref.~\cite{Middleton:2020asl}). While indiscriminately binning in frequency can wash out anisotropies, we showed that a variety of models predict significant values at the highest frequencies accessible to PTAs. This indicates that sensitivity to high frequencies will be crucial to achieve a detection and that the results presented in this work, combined with the sensitivity curve of PTA measurements, can be useful to inform the optimal frequency binning in order to maximize the chance of detection. 

We also investigate the impact of any additional mechanisms driving the binary inspiral on the predicted level of anisotropy, such as interactions with the surrounding stars or gas. The model derived in this work offers an extension of Refs.~\cite{Sesana:2008mz, Sampson:2015ada} to predictions of anisotropy, thereby enabling both the GW spectrum and its anisotropies to be predicted in a self-consistent manner. We explicitly demonstrate that the presence of any additional mechanisms will enhance the detectability of anisotropies relative to the isotropic component, due to a reduction on the number of binaries and attenuation of the spectrum of GWs. While not addressed in this work, alternative theories of gravity which predict scalar and vector dipole radiation will similarly alter the binary inspiral, enhancing the level of anisotropy (see, e.g., Ref.~\cite{Cornish:2017oic}), and can therefore be probed through measurements of anisotropy.

Finally, we use the model presented in this work to derive the expected level of polarization of the SGWB. The background is sourced by a finite number of SMBHBs, which are each inherently polarized as a function of the angle between the binary orbit and the line of sight. Hence, the finite number of binaries will lead to spatial fluctuations in the SGWB, captured by measurements of anisotropy. Although the level of anisotropy in circular polarization has been previously derived in Ref.~\cite{ValbusaDallArmi:2023ydl}, our results provide a connection with the sources in the PTA band and also include the linear polarization component. We explore the impact of a linearly polarized anisotropic background on the PTA signal and its detectability in an upcoming work~\cite{Kumar_prep}.

\acknowledgments
We thank Stephen Taylor for helpful discussions. GSP was supported by the National Science Foundation Graduate Research Fellowship under Grant No.\ DGE1746891. This work was supported at Johns Hopkins by NSF Grant No.\ 1818899 and the Simons Foundation.

\bibliography{ref.bib}
\bibliographystyle{utcaps}

\end{document}